\documentstyle[aps,12pt]{revtex}
  \textheight
23.5cm\topmargin -0.5in \textwidth 15.5cm\oddsidemargin
0.in\evensidemargin 0.in

\begin{document}

\title{Quantization Rules for Bound States of \\
the Schr\"{o}dinger Equation}

\author{Zhong-Qi Ma$^{1),2)}$ \thanks{Electronic address:
mazq@sun.ihep.ac.cn} and Bo-Wei Xu$^{3)}$\thanks{Electronic
address: bwxu@sjtu.edu.cn}}

\address{1) CCAST (World Laboratory), P.O.Box 8730, Beijing 100080, China \\
2) Institute of High Energy Physics, Beijing 100039, China\\
3) Department of Physics, Shanghai Jiaotong University, Shanghai
200030, China}

\maketitle

\date{}

\vspace{5mm}

\begin{abstract}
An exact quantization rule for the bound states of the
one-dimensional Schr\"{o}dinger equation is presented and is
generalized to the three-dimensional Schr\"{o}dinger equation with
a spherically symmetric potential.

\end{abstract}

\section{Introduction}

In the development of quantum mechanics, the Bohr-Sommerfeld
quantization rules of the old quantum theory \cite{sch} occupy a
position intermediate between classical and quantum mechanics. The
WKB approximation \cite{wen,kra,bri} is a method for the
approximate treatment of the Schr\"{o}dinger wave function with
another quantization rule \cite{sch}:
$$\displaystyle \int_{x_{A}}^{x_{B}} k dx =(n+1/2) \pi,~~~~~~
k=\sqrt{2\mu[E-V(x)]}/\hbar,~~~~~~n=0,~1,~2,~\ldots , \eqno (1) $$

\noindent where $x_{A}$ and $x_{B}$ are two turning points
$$E=V(x_{A})=V(x_{B}),~~~~~~EV(x),~~~~~~x_{A}<x<x_{B},
\eqno (2) $$

\noindent and $n$ is the number of nodes of the WKB wave function
between two turning points. The half-integer number is the quantum
correction to the Bohr-Sommerfeld result, which comes from the
fact that the wave function in WKB analysis may extend into the
classically forbidden region. The WKB method is precise for the
harmonic oscillator potential, and is expected to be most useful
in the nearly classical limit of large quantum numbers, namely, it
is good when $n$ is large in comparison with unity. Various
refinements have been developed to improve the accuracy of the WKB
method \cite{xia,pop,ziv} where the main modification was made on
the phase loss at the turning points. Recently, Cao et al.
presented a calculation method, called the analytic transfer
matrix method (ATMM) \cite{cao,zho}, for the energy levels of
bound states of the one-dimensional Schr\"{o}dinger equation,
where the phase contribution picked up by the scattered subwaves
was included in the new quantization rule. The accurate numerical
calculation results for some examples were given there
\cite{cao,zho}. This is a prospective method for calculating the
energy levels of bound states of the Schr\"{o}dinger equation, but
has to be developed and improved further. In this Letter we apply
the matching condition of the logarithmic derivatives (which is a
standard method in quantum mechanics) and the fundamental concept
of the mathematical analysis to the problem of bound states of
one-dimensional Schr\"{o}dinger equation, and derive an exact
quantization rule without any approximation. The quantization rule
is rigorous and general for any Schr\"{o}dinger equation with only
one variable. It is a fundamental problem in quantum mechanics.

The plan of this Letter is as follows. In section 2 we will
develop the method of the numerical calculation for the energy
levels of bound states of the one-dimensional Schr\"{o}dinger
equation by the matching condition of the logarithm derivatives of
the wave function. In section 3 we present an exact quantization
rule for one-dimensional Schr\"{o}dinger equation.  The
quantization rule is proved without any approximation. Its
validity can also be confirmed with some solvable examples. In
section 4 the quantization rule is generalized to the
three-dimensional Schr\"{o}dinger equation with a spherically
symmetric potential. The conclusion is given in section 5.

\section{Matching condition of logarithm derivatives}

Consider the one-dimensional Schr\"{o}dinger equation
$$\displaystyle {d^{2}\over dx^{2}}\psi(x)=-\displaystyle {2\mu\over
\hbar^{2}} \left[E-V(x)\right]\psi(x), \eqno (3) $$

\noindent where $\mu$ is the mass of the particle, and the
potential $V(x)$ is a piecewise continuous real function of $x$.
The logarithm derivative $\phi(x)$ of the wave function $\psi(x)$
is
$$\phi(x)=\displaystyle {1\over \psi(x)}\displaystyle {d\psi(x)\over
dx}. \eqno (4) $$

\noindent From the Schr\"{o}dinger equation (3) we have
$$\displaystyle {d\over dx}\phi(x)=-k(x)^{2}-\phi(x)^{2},~~~~~~E\geq V(x), \eqno (5) $$
$$\displaystyle {d\over
dx}\phi(x)=\kappa(x)^{2}-\phi(x)^{2},~~~~~~E\leq V(x), \eqno (6)
$$

\noindent where
$$\begin{array}{ll}
k(x)=\sqrt{2\mu\left[E-V(x)\right]}/\hbar,~~~~&E\geq V(x),\\
\kappa(x)=\sqrt{2\mu\left[V(x)-E\right]}/\hbar,~~~~&E\leq V(x),\\
k(x)=\kappa(x)=0,~~~~~~~~~~&E=V(x). \end{array}\eqno (7) $$

\noindent It is obvious that $\phi(x)$ decreases monotonically
with respect to $x$ when $E\geq V(x)$, but not monotonically when
$E\leq V(x)$. Note that near a node of the wave function $\psi(x)$
in the region where $E\geq V(x)$, $\phi(x)$ decreases to
$-\infty$, jumps to $+\infty$, and then, decreases again.

Arbitrarily choosing an energy $E$ as a parameter, we assume for
definiteness that the potential $V(x)$ satisfies
$$\begin{array}{ll}
V(x)=V_{I}E, ~~~~~~~~~~&-\infty < x \leq x_{I},\\
V(x)E, ~~~~&x_{I}<x < x_{A}~~{\rm or}~~x_{B}<x<x_{F}, \\
V(x)=E, ~~~~&x=x_{A}~~{\rm or}~~x=x_{B},\\
V(x)<E, ~~~~&x_{A}<x<x_{B},\\
V(x)=V_{F}E ~~~~~~~~~~&x_{F}\leq x < \infty. \end{array} \eqno (8)
$$

\noindent $x_{A}$ and $x_{B}$, $x_{A}<x_{B}$, are called two
turning points where $V(x_{A})=V(x_{B})=E$. This potential can be
easily generalized.

The Schr\"{o}dinger equation (3) is a linear differential equation
of the second order, for which there are two independent
solutions. In the region $-\infty < x \leq x_{I}$, one solution is
divergent exponentially, and the other is physically admissible:
$$\psi(x)\sim e^{\kappa_{I}
x},~~~~~~\phi(x_{I})=\kappa_{I}=\sqrt{2\mu(V_{I}-E)}/\hbar0. \eqno
(9) $$

\noindent Similarly, in the region $x_{F}\leq x <\infty$, the
physically admissible solution is
$$\psi(x)\sim e^{-\kappa_{F}
x},~~~~~~\phi(x_{F})=-\kappa_{F}=-\sqrt{2\mu(V_{F}-E)}/\hbar<0.
\eqno (10) $$

\noindent For the general potential, if $E<V(\pm \infty)$, both
$\phi(x_{I})$ and $-\phi(x_{F})$ are positive .

By making use of the fundamental concept of the mathematical
analysis, we replace the continuous potential well with a stack of
thin films each of which has a constant potential. We first divide
the region $x_{I}\leq x \leq x_{A}$ into $n$ equal films with
width $d_{n}$, where $E\leq V(x)$ and $x_{A}=x_{I}+nd_{n}$. In the
$j$th film, $x_{I}+jd_{n}-d_{n}\leq x \leq x_{I}+jd_{n}$, $V(x)$
is replaced with a constant potential $V_{j}$
$$V_{j}=V(x_{I}+jd_{n}-d_{n}/2),~~~~~~\kappa_{j}=\sqrt{2\mu\left[V_{j}-E\right]}/\hbar.
\eqno (11) $$

\noindent Solving the Schr\"{o}dinger equation at this film, we
obtain
$$\psi_{j}(x)=A_{j}e^{\kappa_{j}x}+B_{j}e^{-\kappa_{j}x} . \eqno (12)
$$

\noindent On two ends of the film, the logarithm derivatives
$\varphi_{j-1}$ and $\varphi_{j}$, which should match with the
logarithm derivatives at the ends of the neighboring films, are
$$\begin{array}{l}
\varphi_{j}=\left. \displaystyle {1\over \psi_{j}(x)}\displaystyle
{d\psi_{j}(x)\over dx}\right|_{x=x_{I}+jd_{n}} =
\kappa_{j}\displaystyle
{A_{j}e^{\kappa_{j}(x_{I}+jd_{n})}-B_{j}e^{-\kappa_{j}(x_{I}+jd_{n})}
\over
A_{j}e^{\kappa_{j}(x_{I}+jd_{n})}+B_{j}e^{-\kappa_{j}(x_{I}+jd_{n})}},\\[2mm]
\varphi_{j-1}=\left. \displaystyle {1\over
\psi_{j}(x)}\displaystyle {d\psi_{j}(x)\over
dx}\right|_{x=x_{I}+jd_{n}-d_{n}} = \kappa_{j}\displaystyle
{A_{j}e^{\kappa_{j}(x_{I}+jd_{n}-d_{n})}-B_{j}e^{-\kappa_{j}(x_{I}+jd_{n}-d_{n})}
\over A_{j}e^{\kappa_{j}(x_{I}+jd_{n}-d_{n})}+B_{j}
e^{-\kappa_{j}(x_{I}+jd_{n}-d_{n})}} . \end{array} \eqno (13) $$

\noindent From the second formula of Eq. (13) we obtain
$$A_{j}e^{\kappa_{j}(x_{I}+jd_{n}-d_{n})}\left\{\kappa_{j}-\varphi_{j-1}\right\}
=B_{j}e^{-\kappa_{j}(x_{I}+jd_{n}-d_{n})}\left\{\kappa_{j}+\varphi_{j-1}\right\}.$$

\noindent Substituting it into the first formula of Eq. (13) we
have
$$\varphi_{j}= \kappa_{j}
\displaystyle {
\kappa_{j}\tanh\left(\kappa_{j}d_{n}\right)+\varphi_{j-1} \over
\kappa_{j}+ \varphi_{j-1}\tanh\left(\kappa_{j}d_{n}\right)}. \eqno
(14)$$

\noindent This is a recursive relation. If $\varphi_{j-1}$ is
positive, $\varphi_{j}$ is positive, too. There is no zero both in
the numerator and in the denominator of Eq. (14), so $\varphi_{j}$
is finite and non-vanishing. Since $\phi(x_{I})$ is positive and
known, one is able to calculate $\varphi_{n}=\phi(x_{A})$ from
$\varphi_{0}=\phi(x_{I}) =\kappa_{I}$ with the recursive relation
(14) as $j$ increases from 1 to $n$. $\phi(x_{A})$ is positive,
finite and non-vanishing (see Appendix A). The calculated
precision depends on the number $n$ of the films. In principle,
one may obtain a precise $\phi(x_{A})$ if $n$ is large enough.

Similar calculation can be made in the region $x_{F}\leq x
<\infty$. The recursive relation (14) becomes
$$\varphi_{j-1}= \kappa_{j}
\displaystyle {
\varphi_{j}-\kappa_{j}\tanh\left(\kappa_{j}d_{n}\right) \over
\kappa_{j}- \varphi_{j}\tanh\left(\kappa_{j}d_{n}\right)}. \eqno
(15)$$

\noindent If $\varphi_{j}$ is negative, $ \varphi_{j-1}$ is
negative, finite and non-vanishing. Since $\phi(x_{F})$ is
negative and known, one is able to calculate
$\varphi_{0}=\phi(x_{B}+)$ from $\varphi_{n}=\phi(x_{F})
=-\kappa_{F}$ with the recursive relation (15) as $j$ decreases
from $n$ to 1. $\phi(x_{B}+)$ is negative, finite and
non-vanishing. In principle, one may obtain a precise
$\phi(x_{B}+)$ if $n$ is large enough. Physically, the fact that
there is no zero of $\phi(x)$ in the regions $-\infty< x \leq
x_{A}$ and $x_{B}\leq x <\infty$ implies that in those two regions
the wave function $\psi(x)$ is exponential decay. However, there
may be a zero of $\phi(x)$ in the other classically forbidden
regions (see the end of this section).

Now, we divide the region $x_{A}\leq x \leq x_{B}$ into $m$ equal
films with width $d_{m}$, where $E\geq V(x)$ and
$x_{B}=x_{A}+md_{m}$. In the $j$th film, $x_{A}+jd_{m}-d_{m}\leq x
\leq x_{A}+jd_{m}$, $V(x)$ is replaced with a constant potential
$V_{j}$
$$V_{j}=V(x_{A}+jd_{m}-d_{m}/2),~~~~~~k_{j}=\sqrt{2\mu\left[E-V_{j}\right]}/\hbar.
\eqno (16) $$

\noindent Solving the Schr\"{o}dinger equation at this film, we
obtain
$$\psi_{j}(x)=C_{j}\sin \left(k_{j}x+\delta_{j}\right) . \eqno (17)
$$

\noindent On two ends of the film, the logarithm derivatives
$\phi_{j-1}$ and $\phi_{j}$, which should match with the logarithm
derivatives at the ends of the neighboring films, are
$$\begin{array}{l}
\phi_{j-1}=\left. \displaystyle {1\over \psi_{j}(x)}\displaystyle
{d\psi_{j}(x)\over dx}\right|_{x=x_{A}+jd_{m}-d_{m}}=k_{j}\cot
\left[k_{j}(x_{A}+jd_{m}-d_{m})+\delta_{j}\right],\\[2mm]
\phi_{j}=\left. \displaystyle {1\over \psi_{j}(x)}\displaystyle
{d\psi_{j}(x)\over dx}\right|_{x=x_{A}+jd_{m}}=k_{j}\cot
\left[k_{j}(x_{A}+jd_{m})+\delta_{j}\right]. \end{array} \eqno
(18)$$

\noindent From Eq. (18) we obtain
$$\phi_{j}=k_{j}\cot\left\{{\rm Arctan}\left(\displaystyle {k_{j}\over
\phi_{j-1}}\right)+k_{j}d_{m}\right\}, \eqno (19) $$

\noindent and
$$\begin{array}{l}
k_{j}d_{m} =-{\rm Arctan}\left(\displaystyle {k_{j}\over
\phi_{j-1}}\right)+{\rm Arctan}\left(\displaystyle {k_{j}\over
\phi_{j}}\right)+q\pi,\\
q=\left\{\begin{array}{ll} 0~~~&{\rm no~zero~of}~\phi(x)~{\rm
occurs~in}~x_{A}+jd_{m}-d_{m}\leq x< x_{A}+jd_{m}\\
1~~~&{\rm a~zero~of}~\phi(x)~{\rm
occurs~in}~x_{A}+jd_{m}-d_{m}\leq x< x_{A}+jd_{m},\end{array}
\right.  \end{array} \eqno (20) $$

\noindent where Arctan $\beta$ denotes the principle value of the
inverse tangent function:
$${\rm Arctan}~ \beta=\alpha,~~~~~~\beta=\tan
\alpha,~~~~~~-\pi/2< \alpha \leq \pi/2. \eqno (21) $$

\noindent Note that $\phi(x)$ decreases monotonically with respect
to $x$ when $E V(x)$ [see Eq. (5)]. If no zero of $\phi(x)$ occurs
in the film $x_{A}+jd_{m}-d_{m}\leq x< x_{A}+jd_{m}$,
$${\rm Arctan}\left(\displaystyle {k_{j}\over \phi_{j-1}}\right)< {\rm
Arctan}\left(\displaystyle {k_{j}\over \phi_{j}}\right)\leq
\pi/2,$$

\noindent Equation (20) holds with $q=0$. If one zero of $\phi(x)$
occurs in the film $x_{A}+jd_{m}-d_{m}\leq x< x_{A}+jd_{m}$,
$\phi_{j-1}\geq 0$, and $\phi_{j}< 0$. Thus,
$${\rm Arctan}\left(\displaystyle
{k_{j}\over \phi_{j-1}}\right)\sim \pi/2,~~~~{\rm and}~~ {\rm
Arctan}\left(\displaystyle {k_{j}\over \phi_{j}}\right)\sim
-\pi/2,$$

\noindent we have to add an additional $\pi$ on the right-hand
side of Eq. (20) such that its right-hand side is positive and
equal to $k_{j}d_{m}$. Since the width $d_{m}$ of the film is very
small, we do not consider the case where more than one zeroes of
$\phi(x)$ occur in the film.

Equation (19) is a recursive relation, with which one is able to
calculate $\phi_{m}=\phi(x_{B}-)$ from $\phi_{0}=\phi(x_{A})$ as
$j$ increases from 1 to $m$. The calculated precision depends on
the number $m$ of the films. In principle, one may obtain a
precise $\phi(x_{B}-)$ if $m$ is large enough.

From the Sturm-Liouville theorem (see Appendix B), as $E$
increases, $\phi(x_{B}-)$ decreases monotonically and
$\phi(x_{B}+)$ increases monotonically. Choosing the parameter $E$
by dichotomy such that $\phi(x_{B}-)$ matches with $\phi(x_{B}+)$,
we obtain a bound state with the energy $E$.

Cao et al. \cite{cao} derived the recursive relations similar to
Eqs. (14), (15) and (19), and calculated some examples with more
precise results than those obtained with the nonintegral Maslov
index and the standard WKB method. Zhou et al. \cite{zho}
calculated a problem with a one-dimensional symmetric double-well
potential. Since the potential is symmetric, $V(-x)=V(x)$, the
solution $\psi(x)$ of the Schr\"{o}dinger equation is even or odd
in the spatial inversion, and then, its logarithm derivative
$\phi(x)$ is odd. If a nontrivial solution $\psi(x)$ is odd,
$\psi(0)=0$ and $\phi(x)$ has to be infinity, $\phi(0+)=+\infty$.
If a nontrivial solution $\psi(x)$ is even, $\psi(0)\neq 0$ and
$\phi(x)$ has to be zero because $\phi(x)$ is odd, namely,
$\phi(0+)=0$. One can calculate the energy levels of this system
by the above method in the half space $0<x <\infty$. The different
boundary conditions of $\phi(0+)$ will give different solutions
with different energies, as calculated in \cite{zho}.

\section{Quantization rule}

In the preceding section, we divided the region $x_{A}\leq x \leq
x_{B}$ into $m$ equal films, where $E\geq V(x)$, and obtained Eq.
(20) for $k_{j}d_{m}$. Summing up Eq. (20) from $j=1$ to $j=m$, we
obtain
$$\begin{array}{l}
\displaystyle \sum_{j=1}^{m}~k_{j}d_{m} =N\pi-{\rm
Arctan}\left(\displaystyle {k_{1}\over \phi_{0}}\right)+{\rm
Arctan}\left(\displaystyle {k_{1}\over \phi_{1}}\right)-{\rm
Arctan}\left(\displaystyle {k_{2}\over \phi_{1}}\right)+{\rm
Arctan}\left(\displaystyle {k_{2}\over
\phi_{2}}\right) \\
~~~-+\ldots-{\rm Arctan}\left(\displaystyle {k_{m-1}\over
\phi_{m-2}}\right)+{\rm Arctan}\left(\displaystyle {k_{m-1}\over
\phi_{m-1}}\right)-{\rm Arctan}\left(\displaystyle {k_{m}\over
\phi_{m-1}}\right)+{\rm Arctan}\left(\displaystyle {k_{m}\over
\phi_{m}}\right). \end{array}  \eqno (22) $$

\noindent where $\phi_{0}=\phi(x_{A})$, $\phi_{m}=\phi(x_{B})$,
and $N$ is the number of zeroes of the logarithm derivative
$\phi(x)$ in the region $x_{A}\leq x< x_{B}$. When $m$ goes to
infinity, $d_{m}$ tends to zero, and the sum in Eq. (22) becomes
an integral. Thus, we obtain a new quantization rule:
$$\begin{array}{l}
\displaystyle \int_{x_{A}}^{x_{B}}k(x)dx =N\pi+\displaystyle
\lim_{m\rightarrow \infty}\left\{-{\rm Arctan}\left(\displaystyle
{k_{1}\over \phi(x_{A})}\right)+{\rm Arctan}\left(\displaystyle{
k_{m}\over \phi(x_{B})}\right)\right. \\
\left.~~~~~~+ \displaystyle \sum_{j=1}^{m-1}~\left[{\rm
Arctan}\left(\displaystyle {k_{j}\over \phi_{j}}\right)-{\rm
Arctan}\left(\displaystyle {k_{j+1}\over
\phi_{j}}\right)\right]\right\},\end{array} \eqno (23) $$

\noindent where $\phi_{j}$ is calculated recursively with Eq.
(19). The first term of the right-hand side of Eq. (23) comes from
the zeroes of the logarithm derivative $\phi(x)$ in the region
$x_{A}\leq x<x_{B}$. Since $\phi(x_{A}) 0$ and $\phi(x_{B})< 0$,
the second and the third terms are vanishing as $m$ goes to
infinity if the potential is continuous at the turning points. The
last sum denotes the phase contribution devoted by the scattered
subwaves.

The formula (23) has another expression. If one changes
$\delta_{j}=\delta^{\prime}_{j}+\pi/2$ in Eq. (17), Eq. (18)
becomes
$$\begin{array}{l}
\phi_{j-1}=-k_{j}\tan
\left[k_{j}(x_{A}+jd_{m}-d_{m})+\delta^{\prime}_{j}\right],\\
\phi_{j}=-k_{j}\tan
\left[k_{j}(x_{A}+jd_{m})+\delta^{\prime}_{j}\right]. \end{array}
$$

\noindent Then, equations (20) and (23) become
$$\begin{array}{l}
k_{j}d_{m} ={\rm Arctan}\left(\displaystyle {\phi_{j-1}\over
k_{j}}\right)-{\rm Arctan}\left(\displaystyle {\phi_{j} \over
k_{j}}\right)+q'\pi,\\
q'=\left\{\begin{array}{ll} 0~~~&{\rm no~zero~of}~\psi(x)~{\rm
occurs~in}~x_{A}+jd_{m}-d_{m}<x\leq x_{A}+jd_{m},\\
1~~~&{\rm a~zero~of}~\psi(x)~{\rm
occurs~in}~x_{A}+jd_{m}-d_{m}<x\leq x_{A}+jd_{m},\end{array}
\right. \end{array} $$
$$\begin{array}{l}
\displaystyle \int_{x_{A}}^{x_{B}}k(x)dx
=N^{\prime}\pi+\displaystyle \lim_{m\rightarrow \infty}\left\{{\rm
Arctan}\left(\displaystyle {\phi(x_{A}) \over k_{1}}\right)-{\rm
Arctan}\left(\displaystyle{\phi(x_{B}) \over
k_{m}}\right)\right.\\
~~~\left.~~~+ \displaystyle \sum_{j=1}^{m-1}~\left[{\rm
Arctan}\left(\displaystyle {\phi_{j} \over k_{j+1} }\right)-{\rm
Arctan}\left(\displaystyle {\phi_{j} \over
k_{j}}\right)\right]\right\},  \end{array} \eqno (24) $$

\noindent where $N^{\prime}$ denotes the number of nodes of the
wave function $\psi(x)$ in the region $x_{A}< x \leq x_{B}$. If
the potential is continuous at the turning points, due to
$\phi(x_{A}) 0$ and $\phi(x_{B})< 0$, the second and the third
terms are
$$\displaystyle \lim_{m\rightarrow \infty}{\rm Arctan}\left(
\displaystyle {\phi(x_{A}) \over k_{1}}\right) -\displaystyle
\lim_{m\rightarrow \infty}{\rm Arctan}\left( \displaystyle
{\phi(x_{B}) \over k_{m}}\right) =\pi/2 -(-\pi/2)=\pi . \eqno (25)
$$

\noindent  Since $\phi(x)$ decreases monotonically in the region
$x_{A}<x<x_{B}$, $N=N^{\prime}+1$.

The sum on the right-hand side of Eq. (23) can be transformed into
an integral expression:
$$\begin{array}{rl}
\displaystyle \int_{x_{A}}^{x_{B}}k(x)dx &=~ N\pi+\displaystyle
\lim_{m\rightarrow \infty}\left\{-{\rm Arctan}\left(\displaystyle
{k_{1}\over \phi(x_{A})}\right)+{\rm Arctan}\left(\displaystyle{
k_{m}\over \phi(x_{B})}\right)\right\}\\[2mm]
&~~~~~~-\displaystyle \int_{x_{A}}^{x_{B}} \displaystyle
{\phi(x)\left(dk(x)/dx\right) \over \phi(x)^{2}+k(x)^{2}} dx.
\end{array} \eqno (26) $$

\noindent Two terms in the curly brackets are vanishing as $m$
goes to infinity if the potential is continuous at the turning
points. The sum in Eq. (24) can also be transformed into an
integral expression.

The quantization rule (26) is proved without any approximation, so
that it is exact. Its validity can also be confirmed by comparing
it with the following solvable examples. Both Eqs. (23) and (24)
are the formulas of numerical calculation for Eq. (26). Cao et al.
\cite{cao,zho} presented an expression similar to Eq. (24) and
demonstrated it to be very effective in numerical calculation
through two examples: the one-dimensional Schr\"{o}dinger equation
with a power-law potential \cite{cao} and a symmetric double-well
potential \cite{zho}.

\vspace{2mm} \noindent {\bf Ex. 1}. The harmonic oscillator
potential.

The WKB method is precise for the harmonic oscillator potential.
Now, we are going to check our new quantization rule (26) for the
harmonic oscillator potential $V(x)=\mu \omega^{2}x^{2}/2$. Let
$$\alpha=\sqrt{\displaystyle {\mu \omega \over
\hbar}},~~~~~~\xi=\alpha x, \eqno (27) $$

\noindent we have
$$\begin{array}{l}\psi_{n}(x)=N_{n}e^{-\xi^{2}/2}H_{n}(\xi),~~~~~~
E_{n}=\hbar \omega\left(n+1/2\right),\\
\phi_{n}(x)=-\alpha \xi+2n \alpha H_{n-1}(\xi)/H_{n}(\xi),
\end{array} \eqno (28) $$

\noindent where $N_{n}$ is the normalization factor, $H_{n}(\xi)$
denote the $n$th Hermitian polynomial. In the region $x_{A}\leq x
\leq x_{B}$, where $-x_{A}=x_{B}=\sqrt{2n+1}/\alpha$, we have
$$k^{(n)}(x)=\alpha \sqrt{2n+1-\xi^{2}},~~~~~~\displaystyle
{dk^{(n)}(x) \over dx}=-\alpha^{4}x/k^{(n)}(x).  \eqno (29) $$

After calculation for $0\leq n \leq 10$, we know that
$\phi_{n}(x_{A})=-\phi_{n}(x_{B})0$, $\phi_{n}(x)$ has $n+1$
zeroes in the region $x_{B}\leq x < x_{B}$, and Eq. (26) becomes
$$\begin{array}{l}\displaystyle \int_{x_{A}}^{x_{B}}k^{(n)}(x)dx
=(n+1)\pi+\Phi_{n},\\[3mm]
\Phi_{n}=-\displaystyle \int_{x_{A}}^{x_{B}} \displaystyle
{\phi(x)\left[dk^{(n)}(x)/dx\right] \over
\phi(x)^{2}+k^{(n)}(x)^{2}} dx\\[3mm]
~~~~~~=2\displaystyle \int_{0}^{\sqrt{2n+1}} \displaystyle
{2nH_{n-1}(\xi)-\xi H_{n}(\xi) \over
\left[2n+1-\xi^{2}\right]H_{n}(\xi)^{2}+\left[2nH_{n-1}(\xi)-\xi
H_{n}(\xi)\right]^{2}} \displaystyle {\xi H_{n}(\xi)d\xi \over
\sqrt{2n+1-\xi^{2}}}\\[3mm]
~~~=2\displaystyle \int_{0}^{\sqrt{2n+1}} \displaystyle
{\left(2n-\xi^{2}\right) d\xi \over \sqrt{2n+1-\xi^{2}}}+F_{n}.
\end{array} \eqno (30) $$

\noindent At least for $0\leq n \leq 10$ we obtain
$\Phi_{n}=-\pi/2$ by Mathematica. In fact, the first term in
$\Phi_{n}$ is $(2n-1)\pi/2$, and through a variable
transformation:
$$\sqrt{2n+1-\xi^{2}}=t\left(\xi+\sqrt{2n+1}\right),~~~~~~
d\xi=-\displaystyle {4\sqrt{2n+1}t dt \over
\left(1+t^{2}\right)^{2}}, $$

\noindent the second integral $F_{n}$ is calculated to be $-n\pi$:
$$\begin{array}{l}F_{0}=0,\\[3mm]
F_{1}=2\displaystyle \int_{0}^{\sqrt{3}}\displaystyle {-2 d\xi
\over \sqrt{3 - \xi^{2}}(1 + \xi^2)}=-2\displaystyle
\int_{0}^{1}\displaystyle {(1+t^{2})dt \over
1-t^{2}+t^{4}}=-\pi,\\[3mm]
F_{2}=2\displaystyle \int_{0}^{\sqrt{5}}\displaystyle
{-4(4\xi^{2}+5) d\xi \over \sqrt{5 -
\xi^{2}}(4\xi^4+4\xi^{2}+5)}\\[3mm]
~~~~~~=-16\displaystyle \int_{0}^{1}\displaystyle {(1+t^{2})(5-6
t^{2}+5t^{4})dt \over 25-76 t^{2}+118 t^{4}-76 t^{6}+25
t^{8}}=-2\pi,\\[3mm]
F_{3}=2\displaystyle \int_{0}^{\sqrt{7}}\displaystyle
{-6(4\xi^{4}+6 \xi^{2}+9) d\xi \over \sqrt{7 -
\xi^{2}}(4\xi^6+9\xi^{2}+9)}\\[3mm]
~~~~~~=-24\displaystyle \int_{0}^{1}\displaystyle
{(1+t^{2})(247-748 t^{2}+1146t^{4}-748 t^{6}+247 t^{8} )dt \over
1444-8052 t^{2}+20652 t^{4}-27512 t^{6}+20652 t^{8}-8052
t^{10}+1444 t^{12}}\\[3mm]
~~~=-3\pi.\end{array} $$

\noindent Thus, we demonstrate that the quantization rule (26) is
the same as Eq. (1) for the harmonic oscillator potential.

\vspace{2mm} \noindent {\bf Ex. 2}. The square well potential.

Discuss a finite square well potential $V(x)$
$$V(x)=\left\{\begin{array}{ll}V_{A} &x\leq -\pi,\\
V_{B} & x\geq \pi,\\
0~~~~&-\pi<x <\pi. \end{array} \right. \eqno (31) $$

\noindent The logarithm derivatives at the turning points
$x_{A}=-\pi$ and $x_{B}=\pi$ are
$$\phi(x_{A})=\kappa_{I}=\sqrt{2\mu(V_{A}-E)}/\hbar,~~~~~~
\phi(x_{B})=-\kappa_{F}=-\sqrt{2\mu(V_{B}-E)}/\hbar, \eqno (32) $$

\noindent when $E<V_{A}$ and $E<V_{B}$. The solution to the
Schr\"{o}dinger equation is
$$\psi_{n}(x)=\sin\left(k^{(n)}x+\delta^{(n)}\right), \eqno (33) $$

\noindent and its logarithm derivative $\phi_{n}(x)$ is
$$\phi_{n}(x)=k^{(n)}\cot\left(k^{(n)}x+\delta^{(n)}\right) . \eqno (34) $$

\noindent $k^{(n)}$ and $\delta^{(n)}$ are determined by the
matching conditions at the turning points $x_{A}$ and $x_{B}$,
$$\tan\left(-k^{(n)}\pi+\delta^{(n)}\right)= k^{(n)}/\kappa_{I},~~~~~~
\tan\left(k^{(n)}\pi+\delta^{(n)}\right)= -k^{(n)}/\kappa_{F}. $$

\noindent Hence, we obtain
$$k^{(n)}=- \left\{{\rm Arctan}\left(k^{(n)}/\kappa_{I}\right) +{\rm
Arctan}\left(k^{(n)}/\kappa_{F}\right)\right\}/(2\pi)+n/2. \eqno
(35) $$

\noindent $\phi_{n}(x)$ has $n$ zeroes in the region $-\pi \leq x
\leq \pi$, and $k(x)$ takes the constant value $k^{(n)}$ in the
region, $dk(x)/dx=0$. The energy levels are $E_{n}=\left(\hbar
k^{(n)}\right)^{2}/(2\mu)$. In terms of Eq. (35), the right-hand
side of Eq. (26) is calculated to be
$$\begin{array}{l}n\pi-{\rm Arctan}\left(\displaystyle {k_{1}\over
\phi(x_{A})}\right)+{\rm Arctan}\left(\displaystyle{ k_{m}\over
\phi(x_{B})}\right)\\
~~~=n\pi-\left\{{\rm Arctan}\left(k^{(n)}/\kappa_{I}\right) +{\rm
Arctan}\left(k^{(n)}/\kappa_{F}\right)\right\}=2\pi k^{(n)}.
\end{array}\eqno (36) $$

\noindent Due to the constant $k(x)$, the left-hand side of Eq.
(26) is equal to the same value:
$$\displaystyle \int_{x_{A}}^{x_{B}}k(x)dx=2\pi k^{(n)}. \eqno (37) $$

\noindent This quantization rule is different from both the
Bohr-Sommerfeld one and that given by the WKB approximation. When
$V_{A}$ and $V_{B}$ tend to infinity, due to Eqs. (32) and (35)
$k^{(n)}$ goes to $n/2$, and the quantization rule (37) coincides
with the Bohr-Sommerfeld one.

\section{Three-dimensional Schr\"{o}dinger equation}

Consider  the three-dimensional Schr\"{o}dinger equation with a
spherically symmetric potential. After separation of the angular
part of the wave function,
$$\psi({\bf r})=r^{-1}R(r)Y^{\ell}_{m}(\theta,\varphi), \eqno (38)
$$

\noindent the radial equation of the Schr\"{o}dinger equation is
$$\displaystyle {d^{2}R(r)\over dr^{2}}=-\displaystyle {2\mu\over
\hbar^{2}}\left\{E-U(r)\right\}R(r),~~~~~~ U(r)= \displaystyle
{\hbar^{2}\ell(\ell+1)\over 2\mu r^{2}}+V(r). \eqno (39) $$

\noindent Since Eq. (39) is similar to Eq. (3), its energy levels
can be calculated by the matching conditions of the logarithm
derivatives, where the logarithm derivative is defined as
$$\phi(r)=R(r)^{-1}\displaystyle {dR(r)\over
dr}. \eqno (40) $$

As an example, we discuss the problem of the hydrogen atom, where
the potential $V(r)$ is
$$V(r)=-\displaystyle {e^{2}\over r}. \eqno (41) $$

\noindent When $r\rightarrow \infty$, we have
$$R(r)\sim \exp\{-r\sqrt{2\mu |E|}/\hbar\},~~~~~~
\phi(r)\sim -\sqrt{2\mu |E|}/\hbar. $$

\noindent When $r\rightarrow 0$, we have
$$R(r)\sim r^{\ell+1}\left[1-\displaystyle {\mu
e^{2}\over \hbar^{2}(\ell+1)}r\right],~~~~~~ \phi(r)\sim
(\ell+1)/r. $$

\noindent By the method of matching condition of logarithm
derivatives, one is able to calculate the energy $E$ with Eqs.
(14), (15) and (19).

On the other hand, if the solution to Eq. (39) has known, we are
able to check whether the quantization rule (26) holds. For the
energy $E_{n}$
$$E_{n}=-\displaystyle {\mu e^{4}\over
2\hbar^{2}n^{2}},~~~~~~n=1,2,3,\ldots, \eqno (42) $$

\noindent we have the solution \cite{sch}
$$R_{n\ell}(r)=N_{n\ell}e^{-\rho/2}\rho^{\ell+1}L_{n+\ell}^{2\ell+1}(\rho),~~~~~~
\rho=\displaystyle {2\mu e^{2}  \over n\hbar^{2}}~r, \eqno (43)$$

\noindent where $N_{n\ell}$ is the normalization factor and
$L_{n+\ell}^{2\ell+1}(\rho)$ is the associated Laguerre
polynomials.

When $\ell 0$, the turning points $r_{A}$ and $r_{B}$ satisfying
$U(r_{A})=U(r_{B})=E_{n}$ are
$$\begin{array}{l}r_{A}=\displaystyle {n\hbar^{2} \over 2\mu
e^{2}}\rho_{A},~~~~~~\rho_{A}= 2\left\{n-\left[n^{2}- \ell(\ell+1)
\right]^{1/2}\right\},\\
r_{B}=\displaystyle {n\hbar^{2} \over 2\mu
e^{2}}\rho_{B},~~~~~~\rho_{B}= 2 \left\{n+\left[n^{2}-
\ell(\ell+1) \right]^{1/2}\right\}. \end{array} \eqno (44) $$

\noindent When $\ell=0$, we define $r_{A}=0$ with
$U(r_{A})=-e^{2}/r_{A}\sim -\infty$. $r_{B}$ with $U(r_{B})=E_{n}$
still satisfies Eq. (37). The momentum $k_{n\ell}(r)$ is
$$k_{n\ell}(r)=\displaystyle {1 \over 2r}\left\{
\left(\rho-\rho_{A}\right)
\left(\rho_{B}-\rho\right)\right\}^{1/2}. \eqno (45) $$

\noindent In the Schr\"{o}dinger equation for the hydrogen atom,
the quantization rule (26) becomes
$$\begin{array}{rl}\displaystyle \int_{r_{A}}^{r_{B}}k_{n\ell}(r)dr
&=~N\pi+\displaystyle \lim_{m\rightarrow \infty}\left\{-{\rm
Arctan}\left(\displaystyle {k_{1}\over
\phi_{n\ell}(r_{A})}\right)+{\rm Arctan}\left(\displaystyle
{k_{m}\over \phi_{n\ell}(r_{B})}\right)\right\}\\[2mm]
&~~~~~~-\displaystyle \int_{r_{A}}^{r_{B}} \displaystyle
{\phi_{n\ell}(r)\left(dk_{n\ell}(r)/dr\right)  \over
\phi_{n\ell}(r)^{2}+k_{n\ell}(r)^{2}} dr. \end{array}  \eqno
(46)$$

\noindent Calculating the left-hand side of Eq. (51), we obtain
with Eq. (50)
$$\displaystyle \int_{r_{A}}^{r_{B}} k_{n\ell}(r) dr=
\displaystyle \int_{\rho_{A}}^{\rho_{B}} \displaystyle {d\rho
\over 2\rho} \left\{\left(\rho-\rho_{A}\right)
\left(\rho_{B}-\rho\right)\right\}^{1/2}=\left[n-\sqrt{\ell(\ell+1)}\right]\pi.
\eqno (47) $$

\noindent Now, we calculate the right-hand side of Eq. (51). The
number $N$ of zeroes of $\phi_{n\ell}(r)$ in the region $r_{A}\leq
r < r_{B}$ is $(n-\ell)$. The logarithm derivatives at the turning
points $r_{A}$ and $r_{B}$ are non-vanishing.  When $\ell=0$,
$r_{A}$ is not the turning point, but $\phi_{n0}(r_{A})=1/r$.
Thus, we have
$$\begin{array}{ll}\displaystyle \lim_{m\rightarrow \infty}{\rm
Arctan}\left(\displaystyle {k_{m}\over
\phi_{n\ell}(r_{B})}\right)=0,~~~~~~&\ell\geq 0,\\
\displaystyle \lim_{m\rightarrow \infty}{\rm Arctan}
\left(\displaystyle {k_{1}\over
\phi_{n\ell}(r_{A})}\right)=0,~~~~~~&\ell0,\\
\displaystyle \lim_{m\rightarrow \infty}{\rm
Arctan}\left(\displaystyle {k_{1}\over
\phi_{n0}(r_{A})}\right)=\displaystyle \lim_{r\rightarrow 0} {\rm
Arctan}\left\{r\sqrt{2\mu
\left(E_{n}+e^{2}/r\right)}/\hbar\right\}=0,~~~~~~&\ell=0.
\end{array} \eqno (48)$$

\noindent Calculating the integral in Eq. (51) by Mathematica, at
least for $n=1$, 2, 3, and $\ell<n$, we obtain:
$$-\displaystyle \int_{r_{A}}^{r_{B}} \displaystyle
{\phi_{n\ell}(r)\left(dk_{n\ell}(r)/dr\right)  \over
\phi_{n\ell}(r)^{2}+k_{n\ell}(r)^{2}} dr=
\left[\ell-\sqrt{\ell(\ell+1)}\right] \pi. \eqno (49)$$

\noindent Therefore, the quantization rule (46) holds for the
hydrogen atom.

\section{Conclusions}

In this Letter, with the matching condition of the logarithm
derivatives and the fundamental concept of the mathematical
analysis, we proposed a formula (23) for numerically calculating
the energy levels of bound states of the Schr\"{o}dinger equation
in one dimension. Calculating the integral form of Eq. (23), we
obtained an exact quantization rule (26) for the bound states of
the one-dimensional Schr\"{o}dinger equation. The exact
quantization rule was generalized to bound states of the
three-dimensional Schr\"{o}dinger equation with a spherically
symmetric potential. The quantization rule was confirmed by
checking some examples where the solutions of the Schr\"{o}dinger
equation are known. Two examples of numerical calculation for the
one-dimensional Schr\"{o}dinger equation with a power-law
potential \cite{cao} and with a symmetric double-well potential
\cite{zho} demonstrated that the exact quantization rule is very
effective in numerical calculation.

Cao et al. \cite{cao,zho} derived a formula similar to Eq. (24).
However, their formulation contained some unclear points. The
integer N in Eq. (29) of Ref. \cite{cao} is unclarified. They
introduced  the ``exponentially decaying coefficients" $P_{j}$,
but the physical meaning of $P_{j}$ is rather ambiguous. In fact,
$P_{j}$ are nothing but the logarithm derivatives of the wave
function (in their region of $0\leq x \leq x_{C}$) or that
multiplied with $-1$ (in their region of $x_{D}\leq x \leq
x_{S}$). This is the reason why the coefficients should be matched
at the turning points, which was not explained clearly in Ref.
\cite{cao}. Finally we would like to point out that the series
form (23) is an approximate formula of the integral form (26)
derived in the present Letter, which is the exact quantization
rule.

\vspace{5mm} \noindent

{\bf ACKNOWLEDGMENTS}. One of the authors (BWX) would like to
thank Professor Z. Cao for drawing his attention to this problem.
This work was supported by the National Natural Science Foundation
of China.

\begin{center}

{\bf Appendix A} $~~$ Property of $\phi(x_{A})$
\end{center}

From the recursive relation (14) and $\phi(x_{I})0$, we have
proved that $\phi(x_{A})$ is non-negative and finite. Now, we are
going to prove that $\phi(x_{A})\neq 0$ by reduction to absurdity.
When $x\leq x_{A}$ and $x$ near $x_{A}$, consider the leading term
in the power series in $(x_{A}-x)$. Due to $\kappa(x_{A})=0$, we
have $\kappa(x)\simeq C(x_{A}-x)^{\alpha}$, where $C0$ and
$\alpha0$. If $\phi(x_{A})=0$, from Eq. (6) we have
$\phi(x_{A})\simeq -C^{2}(x_{A}-x)^{2\alpha+1}/(2\alpha+1)<0$. It
conflicts to the fact that $\phi(x_{A})$ is non-negative.

\vspace{4mm}
\begin{center}

{\bf Appendix B} $~~$ The Sturm-Liouville Theorem

\end{center}

Denote by $\psi(E,x)$ the solution of the Schr\"{o}dinger equation
(3) with the energy $E$. Multiplying Eq. (3) with $\psi(E_{1},x)$
we have
$$\psi(E_{1},x)\displaystyle {\partial ^{2}\over \partial x^{2}}\psi(E,x)
=-\displaystyle {2\mu\over \hbar^{2}}
\left[E-V(x)\right]\psi(E,x)\psi(E_{1},x), \eqno (B1) $$

\noindent Exchanging $E_{1}$ and $E$ and subtracting from it by
Eq. (B1), we obtain
$$\displaystyle {\partial \over
\partial x}\left\{\psi(E,x)\displaystyle {\partial \psi(E_{1},x) \over
\partial x}-\psi(E_{1},x)\displaystyle {\partial \psi(E,x) \over
\partial x}\right\}=-\displaystyle {2\mu \over
\hbar^{2}}\left(E_{1}-E\right) \psi(E,x)\psi(E_{1},x). \eqno (B2)
$$

\noindent When $E<V(-\infty)$, the boundary condition gives that
both solutions $\psi(E,x)$ and $\psi(E_{1},x)$ are vanishing at
negative infinity. Integrating Eq. (B2) from $-\infty$ to
$x_{B}-$, we obtain
$$\begin{array}{l}\displaystyle {1 \over E_{1}-E}\left\{\psi(E,x)\displaystyle {\partial
\psi(E_{1},x) \over \partial x}-\psi(E_{1},x)\displaystyle
{\partial \psi(E,x) \over
\partial x}\right\}_{x=x_{B}-}\\
~~~=-\displaystyle {2\mu \over \hbar^{2}}\displaystyle
\int_{-\infty}^{x_{B}} \psi(E,x)\psi(E_{1},x)dx. \end{array} $$

\noindent Taking the limit as $E_{1}$ goes to $E$, we have
$$\begin{array}{l}\left.\displaystyle {\partial \phi(E,x) \over \partial
E}\right|_{x=x_{B}-} =\displaystyle {\partial \over \partial
E}\left\{ \displaystyle {1 \over \psi(E,x)}\displaystyle {\partial
\psi(E,x) \over \partial x}\right\}_{x=x_{B}-} \\
~~~~~~=-\displaystyle {2\mu \over
\hbar^{2}\psi(E,x_{B})^{2}}\displaystyle \int_{-\infty}^{x_{B}}
\psi(E,x)^{2}dx<0. \end{array} \eqno (B3) $$

\noindent Namely, at a given point $x_{B}-$, the logarithm
derivative $\phi(E,x)$ of the wave function $\psi(E,x)$ decreases
monotonically as $E$ increases.

Similarly, when $E<V(\infty)$, the boundary condition gives that
both solutions $\psi(E,x)$ and $\psi(E_{1},x)$ are vanishing at
positive infinity. Integrating Eq. (B2) from $x_{B}+$ to $\infty$,
we obtain
$$\displaystyle {1 \over E_{1}-E}\left\{\psi(E,x)\displaystyle {\partial
\psi(E_{1},x) \over \partial x}-\psi(E_{1},x)\displaystyle
{\partial \psi(E,x) \over
\partial x}\right\}_{x=x_{B}+}=\displaystyle {2\mu \over
\hbar^{2}}\displaystyle \int_{x_{B}}^{\infty}
\psi(E,x)\psi(E_{1},x)dx.$$

\noindent Taking the limit as $E_{1}$ goes to $E$, we have
$$\left.\displaystyle {\partial \phi(E,x) \over \partial
E}\right|_{x=x_{B}+}  =\displaystyle {2\mu \over
\hbar^{2}\psi(E,x_{B})^{2}}\displaystyle \int^{\infty}_{x_{B}}
\psi(E,x)^{2}dx0. \eqno (B4) $$

\noindent Namely, at a given point $x_{B}+$, the logarithm
derivative $\phi(E,x)$ of the wave function $\psi(E,x)$ increases
monotonically as $E$ increases.


\begin{thebibliography}{99}

\bibitem{sch} L. I. Schiff, Quantum Mechanics, Third Edition,
(McGraw-Hill Book Co., New York, 1968).

\bibitem{wen} G. Wentzel, Z. Physik. {\bf 38}, 518 (1926).

\bibitem{kra} H. A. Kramers, Z. Physik. {\bf 39}, 828 (1926).

\bibitem{bri} L. Brillouin, Compt. Rend. {\bf 183}, 24 (1926).

\bibitem{xia} F. Xiang and G. L. Yip, J. Lightwave Technol. {\bf
12}, 443 (1994).

\bibitem{pop} V. S. Popov, B. M. Karmakov, and V. D. Mur, Phys.
Lett. A {\bf 210}, 402 (1996).

\bibitem{ziv} S. Zivanovic, V. Milanovic, and Z. Ikonic, Phys.
Status Solidi B {\bf 204}, 713 (1997).

\bibitem{cao} Z. Cao, Q. Liu, Q. Shen, X. Dou, Y. Chen, and Y. Ozaki,
Phys. Rev. A {\bf 63}, 054103 (2001).

\bibitem{zho} F. Zhou, Z. Cao, and Q. Shen, Phys. Rev. A {\bf 67}, 062112 (2003).


\end{thebibliography}
\end{document}